\documentclass[12pt,a4paper,twoside]{article}
\usepackage{graphics}
\usepackage{delarray}
\newlength{\figwidth}
\newlength{\figheight}
\newcommand{\eqref}[1]{(\ref{#1})}
\title{
\vspace{-1.5cm}
\begin{flushright}
{\normalsize LC-PHSM-2001-020} \\[-0.5cm]
{\normalsize February 20, 2001}
\end{flushright}
\vspace{1cm}
Measurement of the leptoquark Yukawa couplings \\ 
       in $e^{+}e^{-}$ collisions at TESLA}
\author{Aleksander Filip \.Zarnecki \\
{\small\it Institute of Experimental Physics, Warsaw University,} \\
{\small\it Ho\.za 69, 00-681 Warszawa, Poland} \\
{\small\it E-mail: zarnecki@fuw.edu.pl}
} 
\date{} 
\begin{document} 

\maketitle 

\begin{abstract}

Measurement of the Yukawa couplings of the first-generation leptoquarks
has been studied for $e^{+}e^{-}$ collisions at TESLA, at $\sqrt{s}=$800 GeV.
By combining measurements from different production and decay channels,
determination of Yukawa couplings with precision on the few per-cent level
is possible.
TESLA will be sensitive to very small leptoquark Yukawa couplings
not accessible at LHC, down to $\lambda_{L,R} \sim 0.05 e$.
Distinction between left-handed and right-handed Yukawa 
couplings is feasible even for leptoquark masses very close to the
pair-production kinematic limit.

\end{abstract}

\thispagestyle{empty}

%
%

\section{Introduction}
\label{sec-intro}

New result on atomic parity violation (APV) in Cesium and 
unitarity of the CKM matrix, as well as recent LEP2 hadronic 
cross-section measurements indicate possible 
deviations from the Standard Model predictions. 
Exchange of leptoquark type objects with masses above 250GeV 
has been proposed as a possible explanation for these effects \cite{myglqa}. 
If the leptoquark Yukawa couplings are assumed to be small, 
leptoquark masses of the order of 300 GeV are consistent 
with the exiting data.

Running at high luminosity LHC experiments will search for 
leptoquark pair production for leptoquark masses up to 
about 1.5 TeV \cite{cmslq,myfuture}.
Leptoquark mass can be determined at LHC with high precision, limited
mainly by the systematic uncertainty of the energy scale.
However, leptoquark pair production in strong interactions is
not sensitive to the leptoquark Yukawa couplings.
%
%
Yukawa couplings could be measured in the Drell-Yan process 
at LHC for coupling to the mass ratios $\lambda / M$ 
down to about 0.1 TeV$^{-1}$ \cite{myfuture}.
However, the measurement is only possible if the leptoquark type 
(spin, corresponding quark flavor, coupling chirality) is 
known.\footnote{For different leptoquark models contribution to the 
Drell-Yan lepton pair production cross-section can differ even 
by an order of magnitude.}
The value of the leptoquark pair production cross-section 
and the leptoquark branching fractions will allow differentiation  
between some leptoquark models at LHC.
But the precise determination of the leptoquark type may not be possible.

For masses below 400 GeV leptoquarks can be also pair produced in
$e^{+}e^{-}$ collisions at TESLA, at $\sqrt{s}=800$ GeV.
The value of the total cross-section and the angular 
distribution of the produced leptoquark pairs depend
on the leptoquark type and the leptoquark Yukawa couplings.
Also the single leptoquark production can be used to constrain
leptoquark parameters.

In this note, precision with which leptoquark Yukawa couplings
can be determined at TESLA is studied. 
The leptoquark models used in this analysis are described 
in  section \ref{sec-model}. 
Results from the global analysis of available experimental data \cite{myglqa}
and the possible leptoquark signal are briefly summarized in
section \ref{sec-cur}.
Leptoquark production at TESLA is described in section \ref{sec-prod}
and the event selection methods in section \ref{sec-sel}.
Yukawa coupling determination from the observed angular distributions
is described in section \ref{sec-res}.


\section{Leptoquark models}
\label{sec-model}

In this paper a general classification of leptoquark states 
proposed by Buchm\"uller, R\"uckl and Wyler \cite{brw} will be used.
The Buchm\"uller-R\"uckl-Wyler (BRW) model is based on 
the assumption that new interactions should respect the 
$SU(3)_{C} \times SU(2)_{L} \times U(1)_{Y}$ symmetry of
the Standard Model.
In addition leptoquark couplings are assumed to be family diagonal
(to avoid FCNC processes) and to conserve lepton and baryon numbers
(to avoid rapid proton decay).
%
%
With all these assumptions there are 10 possible states 
(isospin singlets or multiplets) of scalar and vector leptoquarks.
Table \ref{tab-aachen} lists these states according to 
the so-called Aachen notation \cite{aachen}.
An S(V) denotes a scalar(vector) leptoquark and the subscript
denotes the weak isospin.
A tilde is introduced to differentiate between leptoquarks
with different hypercharges.
\begin{table}[tbp]
  \begin{center}
   \begin{tabular}{lcccccc}
      \hline\hline\hline\noalign{\smallskip}
Model & Fermion & Charge & $BR(LQ \rightarrow e^{\pm}q)$ & 
       $LQ$-$e$-$q$ &  \multicolumn{2}{c}{Channel}  \\
      & number F &   Q   & $\beta$  &  Coupling &  &  \\
\hline\hline\hline\noalign{\smallskip}
$S_{\circ}$ &  2  &  $-1/3$  &  $\frac{1+r}{2+r}$  &  $\lambda_{L}$ & $e u$ & $\nu d$  \\
            &     &          &                     &  $\lambda_{R}$ & $e u$ &          \\
\hline\noalign{\smallskip}
$\tilde{S}_{\circ}$  &  2  &  $-4/3$   &  1  &  $\lambda_{R}$ & $ed$ &     \\
\hline\noalign{\smallskip}
$S_{1/2}$   &  0  &  $-5/3$   &  1  &  $\lambda_{L}$  & $e \bar{u}$ &   \\
            &     &           &     &  $\lambda_{R}$  & $e \bar{u}$ &   \\
\noalign{\smallskip}
            &     &  $-2/3$   &  $\frac{1}{1+r}$ &  $\lambda_{L}$  &   & $\nu \bar{u}$  \\
            &     &           &                  &  $\lambda_{R}$  & $e \bar{d}$ &  \\
\hline\noalign{\smallskip}
$\tilde{S}_{1/2}$ &  0 &  $-2/3$  &  1  &  $\lambda_{L}$  &   $e \bar{d}$  &   \\
                  &    &  $+1/3$  &  0  &  $\lambda_{L}$  &   & $\nu \bar{d}$  \\
\hline\noalign{\smallskip}
$S_{1}$       &  2  & $-4/3$  &  1  & $\sqrt{2} \lambda_{L}$  &   $ed$ &  \\
              &     & $-1/3$  &  1/2 & $\lambda_{L}$  &   $eu$ &  $\nu d$   \\
              &     & $+2/3$  &  0   &  $\sqrt{2} \lambda_{L}$  &   &  $\nu u$    \\
\hline\hline\hline\noalign{\smallskip}
$V_{\circ}$ &  0  &  $-2/3$& $\frac{1+r}{2+r}$ & $\lambda_{L}$ &$e\bar{d}$&$\nu \bar{u}$ \\
            &     &          &               &  $\lambda_{R}$ & $e\bar{d}$ &          \\
\hline\noalign{\smallskip}
$\tilde{V}_{\circ}$  &  0  &  $-5/3$   &  1  &  $\lambda_{R}$ & $e\bar{u}$ &     \\
\hline\noalign{\smallskip}
$V_{1/2}$   &  2  &  $-4/3$   &  1  &  $\lambda_{L}$  & $ed$ &   \\
            &     &           &     &  $\lambda_{R}$  & $ed$ &   \\
\noalign{\smallskip}
            &     &  $-1/3$   &  $\frac{1}{1+r}$ &  $\lambda_{L}$  &   & $\nu d$  \\
            &     &           &                  &  $\lambda_{R}$  & $e u $ &  \\
\hline\noalign{\smallskip}
$\tilde{V}_{1/2}$ &  2 &  $-1/3$  &  1  &  $\lambda_{L}$  &   $e u$  &   \\
                  &    &  $+2/3$  &  0  &  $\lambda_{L}$  &   & $\nu u$  \\
\hline\noalign{\smallskip}
$V_{1}$       &  0  & $-5/3$  &  1  & $\sqrt{2} \lambda_{L}$  &   $e\bar{u}$ &  \\
              &     & $-2/3$  &  1/2 & $\lambda_{L}$  &   $e \bar{d}$ &  $\nu \bar{u}$   \\
              &     & $+1/3$  &  0   &  $\sqrt{2} \lambda_{L}$  &   &  $\nu \bar{d}$    \\
\hline\hline\hline\noalign{\smallskip}
    \end{tabular}
  \end{center}
  \caption{A general classification of leptoquark states 
 in the Buchm\"uller-R\"uckl-Wyler model. 
 Listed are the leptoquark fermion number, F, 
 electric charge, Q (in units of elementary charge), 
the branching ratio to electron-quark (or electron-antiquark), $\beta$,
leptoquark-electron-quark couplings in terms of the Yukawa couplings $\lambda_{L}$ and
$\lambda_{R}$,  and the flavors of the coupled lepton-quark pairs. 
For $S_{\circ}$, $^{-2/3}S_{1/2}$, $V_{\circ}$ and $^{-1/3}V_{1/2}$  leptoquarks, 
the branching ratio $\beta$ depends on the coupling ratio 
$r=\lambda_{R}^{2}/\lambda_{L}^{2}$.
}
  \label{tab-aachen}
\end{table}
Listed in Table \ref{tab-aachen} are the leptoquark fermion
number F, electric charge Q, and the branching ratio to an electron-quark
pair (or electron-antiquark pair), $\beta$.
%
%
For a given electron-quark branching ratio $\beta$, the branching ratio 
to the neutrino-quark is by definition $(1-\beta)$. 
Also included in Table \ref{tab-aachen} are leptoquark-electron-quark couplings,
in units of the Yukawa couplings $\lambda_{L}$ and $\lambda_{R}$, 
and the flavors of the lepton-quark pairs coupling to a given leptoquark type.
Strong bounds from rare decays \cite{brw,pion} 
indicate that leptoquarks couple only
either to left- or to right-handed leptons, 
i.e. $\lambda_{L} \cdot \lambda_{R} = 0$.
With this additional constraint, four leptoquark models 
($S_{\circ}$, $S_{1/2}$, $V_{\circ}$ and $V_{1/2}$) have to be considered
separately for left- and right-handed couplings (assuming $\lambda_{R}=0$ and 
$\lambda_{L}=0$ respectively). 
In that case, an additional superscript indicates  the coupling chirality.
Present analysis takes into account only leptoquarks which couple
to the first-generation leptons ($e$, $\nu_{e}$) and first-generation 
quarks ($u$, $d$), as they can be best studied at TESLA.
Second- and third-generation leptoquarks as well as generation-mixing
leptoquarks will not be considered in this paper.
It is also assumed that one of the leptoquark types gives the dominant 
contribution, as compared with other leptoquark states 
and that the interference between different leptoquark states can be neglected.
Using this simplifying assumption, different leptoquark types 
can be considered separately.
Finally, it is assumed that different leptoquark states within 
isospin doublets and triplets have the same mass.


\section{Current limits from global analysis}
\label{sec-cur}

In a recent paper\cite{myglqa} available data from HERA, LEP 
and the Tevatron, as well as from low energy  
experiments are used to constrain the Yukawa couplings $\lambda$
and masses $M$ for scalar and vector leptoquarks.
To compare the data with predictions of the BRW model
the global probability  function ${\cal P}(\lambda,M)$ 
is introduced, describing the probability that the data come from
the model described by parameters $\lambda$ and $M$. 
The probability  function  is defined in such a way that the Standard Model 
probability ${\cal P}_{SM} \equiv 1$.
Constraints on the leptoquark couplings and masses were
studied in the limit of very high leptoquark masses
(using the contact interaction approximation \cite{lqci})
as well as for finite leptoquark masses, with mass effects 
correctly taken into account.
Excluded on  95\% confidence level are all models (parameter values) 
which result in  the global probability
less than 5\% of the Standard Model probability:
${\cal P}(\lambda,M) < 0.05$.
For models which describe the data
much better than the Standard Model
(${\displaystyle {\cal P}_{max} \; \equiv \;
\max_{\lambda,M} {\cal P}(\lambda,M) \gg 1}$)
the 95\% CL  signal limit is defined by the condition:
${\cal P}(\lambda,M) > 0.05 \cdot {\cal P}_{max}$.

Current limits on the leptoquark masses are mainly based on the negative
search results at the Tevatron. For scalar leptoquark models considered in 
this paper 95\% CL exclusion limits on the leptoquark masses are 
between 213 and 245 GeV.
Data sensitive to the virtual leptoquark exchange result in limits
on the leptoquark mass $M$ to the Yukawa coupling $\lambda$.
95\% CL exclusion limits on $M/\lambda$, obtained from the global analysis
of existing data, range for the scalar leptoquarks from about 2 to 4 TeV.
Assuming leptoquark mass of 350 GeV, this corresponds to the limits on
the scalar leptoquark Yukawa coupling between 0.09 and 0.17 
( 0.27$e$ to 0.53$e$ ; $e=\sqrt{4\pi\alpha_{em}}$).

\begin{figure}[t]
\centerline{\resizebox{!}{10cm}{%
  \includegraphics{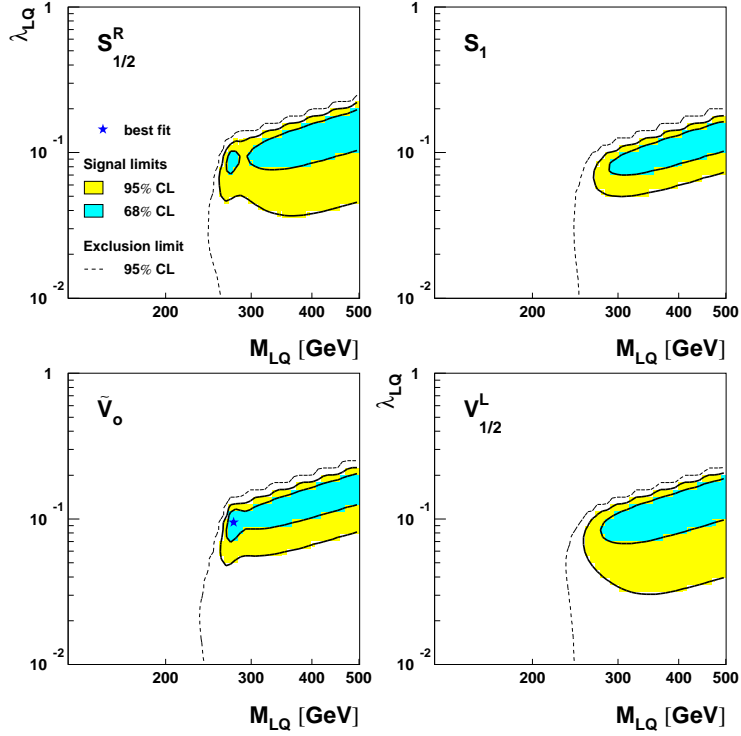}
}}
  \caption{ Signal limits on 68\% and 95\% CL for different leptoquark
            models (as indicated in the plot) resulting from the
            global analysis of existing data \cite{myglqa}.
            Dashed lines indicate the 95\% CL exclusion limits. 
            For $\tilde{V}_{\circ}$ model
            a star indicates the best fit parameters.
            For other models the best fit is obtained in the contact
            interaction limit $M_{LQ} \rightarrow \infty$.
            For $S_{1/2}$ and $V_{1/2}$ models, the superscript indicates
            the considered chirality of the Yukawa coupling.}
  \label{fig-lqsig}
\end{figure}

Four leptoquark models are found to describe the existing experimental data 
much better than the Standard Model 
(${\cal P}(\lambda,M)>20$).
The signal limits for these models, 
at 68\% and 95\% CL are compared with exclusion limits  
in the $(\lambda,M)$ space in Figure \ref{fig-lqsig}.
For $S_{1}$ and $\tilde{V}_{\circ}$ leptoquarks
the observed increase in the global probability 
by factor 367 and 142 respectively
corresponds to more than a 3$\sigma$ effect.
The leptoquark ``signal'' is mostly resulting from the new data 
on the atomic parity violation (APV) in cesium\cite{apvnew}.
After the theoretical uncertainties have been significantly reduced,
the measured value of the cesium weak charge is 
now 2.5$\sigma$ away from the Standard Model prediction.
Also the new hadronic cross-section measurements at LEP2,
for $\sqrt{s}$=192--202 GeV, are on average about 2.5\% above
the predictions\cite{lepnew}.
The effect is furthermore supported by the slight violation of
the CKM matrix unitarity and  HERA high-$Q^{2}$ results.
Existence of leptoquark state with mass of 300--400 GeV
and Yukawa coupling of the order of 0.3$e$ could explain these effects.


\section{Leptoquark production at TESLA}
\label{sec-prod}

Taking into account existing experimental constraints it is rather
unlikely that leptoquarks have masses below 250 GeV, so they can be pair 
produced at TESLA at $\sqrt{s}=500$ GeV.
Therefore, only the high energy TESLA running option is considered 
in this paper, with  $\sqrt{s}=800$ GeV and expected integrated 
luminosity ${\cal L} = 500$ fb$^{-1}$.
%

\begin{figure}[tbp]
\centerline{\resizebox{!}{3cm}{%
  \includegraphics{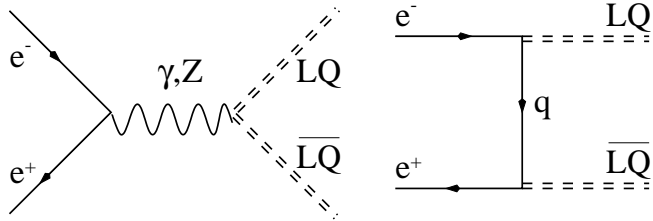}
}}
  \caption{Diagrams for leptoquark pair production in $e^{+}e^{-}$ collisions.}
  \label{fig-diag2}
\end{figure}

In $e^{+}e^{-}$ annihilation, leptoquarks can be pair-produced via photon
and $Z$ boson $s$-channel exchange and via $t$-channel exchange 
of quarks, as shown in Figure \ref{fig-diag2}.
Contribution from $\gamma$ and $Z$ exchange is determined by the leptoquark 
type: its spin, charge and weak isospin.
Only the process with the  $t$-channel quark  exchange is sensitive to the
leptoquark-electron-quark Yukawa coupling.
The value of the Yukawa coupling influences both the total pair production
cross-section value\footnote{Because of the interference with the  $s$-channel 
$\gamma$ and $Z$ exchange,  $t$-channel quark  exchange can increase 
or decrease the total leptoquark pair production cross-section.}
as well as the angular distribution of the produced leptoquark pairs.
Analysis of the leptoquark pair production presented in this paper 
is based on the formulae and prescriptions given in \cite{lqpair_old}
and implemented in the LQPAIR program \cite{lqpair_prog}%
\footnote{Generation of the leptoquark decay angles in LQPAIR 
          has been corrected.}. 

\begin{figure}[tbp]
\centerline{\resizebox{!}{3cm}{%
  \includegraphics{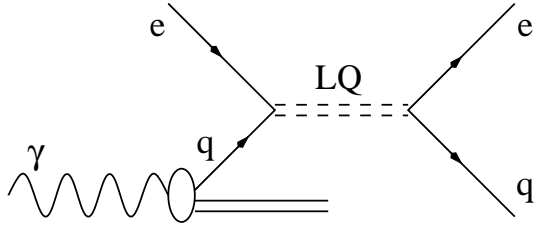}
}}
  \caption{Diagram for single leptoquark production in 
                             $e^{\pm} \gamma $ collisions.}
  \label{fig-diag}
\end{figure}

High energy $e^{+}e^{-}$ collisions can be also used
to study single leptoquark production in $e^{\pm}\gamma$
collisions.  
Considered in this paper are $e^{\pm}\gamma$ collisions 
resulting from the effective electron and positron photon flux, 
described by the  Weizs{\"a}cker-Williams Approximation (WWA), 
and from the electron and positron beam beamstrahlung, 
as simulated by  CIRCE \cite{circe}.
It is assumed that the leptoquark is produced in the electron fusion 
with a quark inside the photon,\footnote{An alternative approach is
to consider direct process $e\gamma \rightarrow LQ \; q$ \cite{egdir}. 
Both approaches give very similar results \cite{myfuture}.}
as shown in Figure \ref{fig-diag} and implemented in PYTHIA \cite{pythia}.


In addition to the leptoquark production, following background processes 
has been generated with PYTHIA version 6.152:
\begin{itemize}
\item Z-pair production: $e^{+}e^{-} \rightarrow Z^{\circ} Z^{\circ}$,
\item W-pair production: $e^{+}e^{-} \rightarrow W^{+} W^{-}$,
\item top quark pair production: $e^{+}e^{-} \rightarrow t \bar{t}$,
\item other (``light'') fermion pair production: 
            $e^{+}e^{-} \rightarrow f \bar{f}$,
      ($f \ne t$),
\item Neutral Current $e^{\pm} \gamma$ deep inelastic scattering (NC DIS):
           $e^{\pm} \gamma \rightarrow e^{\pm} X $.     
\end{itemize}
Cross-section values for selected signal and background samples, 
numbers of generated Monte Carlo events and effective Monte Carlo 
luminosities are summarized in Table \ref{tab-mc}.
When generating signal and background events,
radiative effects were taken into account in both LQPAIR and PYTHIA.
Incoming beams energies were smeared using CIRCE package 
(version 5, revision 1998 05 05) \cite{circe}. 
Parton showering and hadronisation was done with JETSET.
Detector response was simulated using a parametric Monte Carlo
program SIMDET (version 3.01) \cite{simdet}.
Energy flow algorithm implemented in SIMDET was used to join tracker 
and calorimeter information.
Durham jet finding algorithm was then used to group energy flow objects
(corresponding, in most cases, to single particles) into given number
of clusters. 

\begin{table}[t]
  \begin{center}
   \begin{tabular}{lcrr}
\hline\hline\hline\noalign{\smallskip}
Generated &  $\sigma_{e^{+}e^{-}}$ & Generated & Luminosity \\
process &  [fb] & events & [fb$^{-1}$]\\
\hline\hline\hline\noalign{\smallskip}
$e^{+}e^{-} \rightarrow ^{-5/3}S_{1/2}\; ^{+5/3}\bar{S}_{1/2}$ & 
27.0 & 13485 & 500 \\
$e^{+}e^{-} \rightarrow ^{-2/3}S_{1/2}\; ^{+2/3}\bar{S}_{1/2}$ & 
 11.6 & 28963 & 2500 \\
$e^{+}e^{-} \rightarrow ^{-1/3}S_{\circ}\; ^{+1/3}\bar{S}_{\circ}$ & 
 1.11 & 11110 & 10000 \\
$e^{+}e^{-} \rightarrow ^{-4/3}S_{1}\; ^{+4/3}\bar{S}_{1}$ & 
 28.2 & 14100 & 500 \\
$e^{+}e^{-} \rightarrow ^{-1/3}S_{1}\; ^{+1/3}\bar{S}_{1}$ & 
 1.11 & 11110 & 10000 \\
$e^{\pm} \gamma \;\; \rightarrow ^{\pm 1/3}S_{\circ} \; X $ &
                           1.92 & 10000 & 5200 \\
\hline\noalign{\smallskip}
$e^{+}e^{-} \rightarrow Z^{\circ} Z^{\circ}$ & 313 & 240000 & 766 \\
$e^{+}e^{-} \rightarrow W^{+} W^{-}$ & 4320 & 657000 & 152 \\
$e^{+}e^{-} \rightarrow t \bar{t}$  & 306 & 192000 & 628 \\
$e^{+}e^{-} \rightarrow f \bar{f}$ ($f \ne t$) & 6860 & 960000 & 140 \\
$e^{\pm} \gamma \;\; \rightarrow e^{\pm} X $ &
                           2060 & 400000 & 194 \\
\hline\hline\hline\noalign{\smallskip}
    \end{tabular}
  \end{center}
  \caption{Cross-section values, numbers of generated Monte Carlo events
and effective Monte Carlo luminosities, for different processes considered
in this paper. All background samples used in the analysis are shown, 
but only selected signal samples. Signal samples were generated
for leptoquark mass $M$=350 GeV, 
$\lambda_{L} = \lambda_{R} = 0$ for pair production or 
$\lambda_{L} = 0$ and $\lambda_{R} = 0.1\;e$ for single leptoquark production.
Background from $e^{\pm}\gamma$ NC DIS was generated with cut on 
electron-quark invariant mass $M_{eq} > 300$ GeV. 
Cross-sections are given for $e^{+}e^{-}$ scattering 
at $\sqrt{s}=800$ GeV, including beamstrahlung and radiative corrections.}
  \label{tab-mc}
\end{table}
%


\section{Event selection}
\label{sec-sel}

Three leptoquark production and decay channels were considered
in this analysis:
\begin{itemize}
  \item  pair-production, 
  with both leptoquarks decaying to $e^{\pm}$ and jet ($eejj$ events),
  \item  pair-production, 
with one leptoquark decaying to $e^{\pm}$  and jet 
and the other one to neutrino and jet ($e\nu jj$ events),
  \item single-production, with leptoquark decaying to $e^{\pm}$ and jet
             ($ej(j)$ events). 
\end{itemize}
For all channels, event selection cuts were introduced to suppress
background from other processes and allow good event reconstruction.
Selection algorithms were optimized for reconstruction of leptoquark
production events with $M_{LQ}$=350 GeV.

\subsection{Selection of $eejj$ events}

    Cuts used to select leptoquark pair-production candidate events, 
with both leptoquarks decaying into electron (positron) and jet:
\begin{eqnarray*}
e^{+}e^{-} \; \rightarrow & LQ \; \overline{LQ} & \rightarrow \; ee jj
\end{eqnarray*}
    
\begin{itemize}

\item Pre-selection cuts:
      \begin{itemize}
      \item total energy $E$ of an event: $E > 0.9 \sqrt{s}$,
      \item total transverse energy $E_{t}$ of an event: 
                               $E_{t} > 0.4 \sqrt{s}$,
      \item total transverse momentum $p_{t}$ to $E_{t}$ ratio:
                              $p_{t}/E_{t} < 0.1$,
      \end{itemize}

\item Four energy clusters,
      with transverse momenta $p_{t}^{i} > 20$ GeV,
      are reconstructed using the Durham jet algorithm
      ($y_{cut}$ not constrained):
      \begin{itemize}
      \item two clusters are identified as an isolated electron and positron,
      \item two clusters are identified as high multiplicity 
                         ($N_{p}\ge5$) jets.
      \end{itemize}

\item Z-pair background rejection. \\
  Rejected are events fulfilling any of the following conditions:
      \begin{itemize}    
      \item invariant mass of the $e^{+}e^{-}$ system $80 < M_{ee} < 100$ GeV,
      \item invariant mass of the two jet system $70 < M_{jj} < 100$ GeV,
      \item sum of two masses  $M_{ee}+ M_{jj} < 200$ GeV.
      \end{itemize}

\item difference between reconstructed masses of leptoquark ($e^{-}$-jet)
      and anti-leptoquark ($e^{+}$-jet): $\Delta M_{ej} < 60$ GeV,

\item difference between leptoquark and anti-leptoquark 
      azimuthal angle\footnote{Emission angle measured in 
     the plane perpendicular  to the beam axis.}:
       $| \Delta \phi - \pi | \! < \! 0.15$,   
    
\item reconstructed mass within $\pm\; 25$ GeV from the nominal value.
    
\end{itemize}

\begin{figure}[tbp]
\centerline{\resizebox{!}{\figheight}{%
  \includegraphics{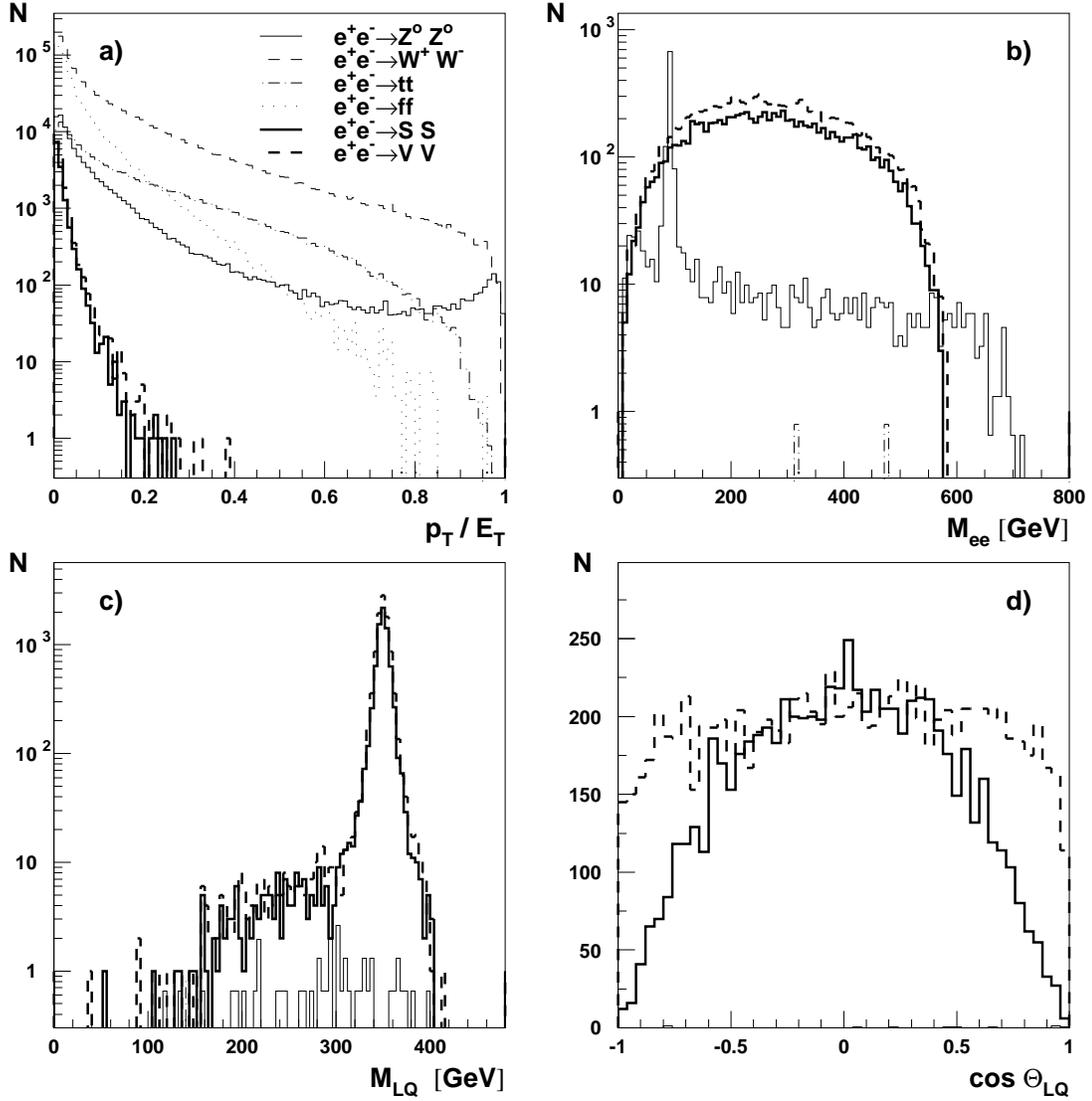}
}}
  \caption{Selection of the leptoquark pair-production events
           in the $eejj$ channel. Distributions for signal 
           (scalar $S_{1/2}$ or vector $V_{\circ}$
            leptoquarks with M=350 GeV) and background events 
           ($Z^{\circ}$ pair-production, $W^{\pm}$ pair-production,
            $t \bar{t}$ pair-production and production of other
            fermion pairs $f \bar{f}$ ($f \ne t$), as indicated in the plot),
            expected for the integrated luminosity of 500$fb^{-1}$, 
            are shown for:
ratio of the total transverse momentum to the transverse energy, before
any selection cuts (a), 
invariant mass  of the identified $e^{+}e^{-}$ pair $M_{ee}$, 
after pre-selection and lepton identification cuts (b),
reconstructed leptoquark mass $M_{LQ}$, after all selection cuts (c),
leptoquark production angle $\Theta_{LQ}$, in $\pm\;25$ GeV mass window (d).
}
  \label{fig-cuts1}
\end{figure}

Expected signal and background event distributions in the main selection 
variables and in the reconstructed leptoquark production angle
are presented in Figure \ref{fig-cuts1}.
The selection efficiency for scalar and vector leptoquark
pair-production, for leptoquark mass $M_{LQ} = 350$ GeV and
$\sqrt{s} = 800$ GeV, is about 50\%. 
Estimated background from Z-pair production\footnote{No other
background Monte Carlo events survive the cuts}
corresponds to the cross-section of about 0.015 $fb$
(7 events in 500 $fb^{-1}$) and is considered to be negligible.

\subsection{Selection of $e\nu jj$ events}

    Main cuts used to select leptoquark pair-production candidate events, with
one leptoquark decaying to electron (positron) and jet and the other
one to neutrino and jet:
\begin{eqnarray*}
e^{+}e^{-} \; \rightarrow &  LQ \; \overline{LQ} & \rightarrow \; e\nu jj
\end{eqnarray*}
    
\begin{itemize}

\item Pre-selection cuts:
      \begin{itemize}
      \item total transverse energy $E_{t}$ of an event: 
                               $E_{t} > 0.25 \sqrt{s}$,      
      \item total transverse momentum $p_{t}$ to $E_{t}$ ratio:
                              $p_{t}/E_{t} > 0.1$,
      \item total invariant mass  $M_{tot}$ of  an event:  
                       $M_{tot} > 0.5 \sqrt{s}$,             
%
%
      \end{itemize}

\item Three energy clusters,
      with energy $100 < E_{i} < 300$ GeV and
      transverse momenta $p_{t}^{i} > 50$ GeV,
      are reconstructed using the Durham jet algorithm, with $y_{cut}$=0.006:
      \begin{itemize}
      \item one cluster is identified as an isolated electron or positron,
      \item two clusters are identified as high multiplicity ($N_p\ge5$) jets.
      \end{itemize}

\item W-pair background rejection.\\
      Rejected are events fulfilling any of the following conditions:
      \begin{itemize}    
      \item invariant mass of the two jet system $M_{jj} < 110$ GeV,
      \item reconstructed $e \nu$ invariant mass $M_{e\nu} < 150$ GeV,
      \item sum of two masses  $M_{jj} + M_{e\nu} < 350$,
      \end{itemize}

\item difference between mass of the (anti-)leptoquark reconstructed 
      from electron-jet and neutrino-jet final state,
       $-80 < M_{ej} - M_{\nu j} < 50$ GeV,

\item  lepton emission angle in the (anti-)leptoquark rest frame
       $\cos \theta^{\star}_{l} > -0.7$,   
    
\item  mass reconstructed from electron-jet final state $M_{ej}$
       within $\pm 20$ GeV from the nominal value.
    
\end{itemize}

\begin{figure}[tbp]
\centerline{\resizebox{!}{\figheight}{%
  \includegraphics{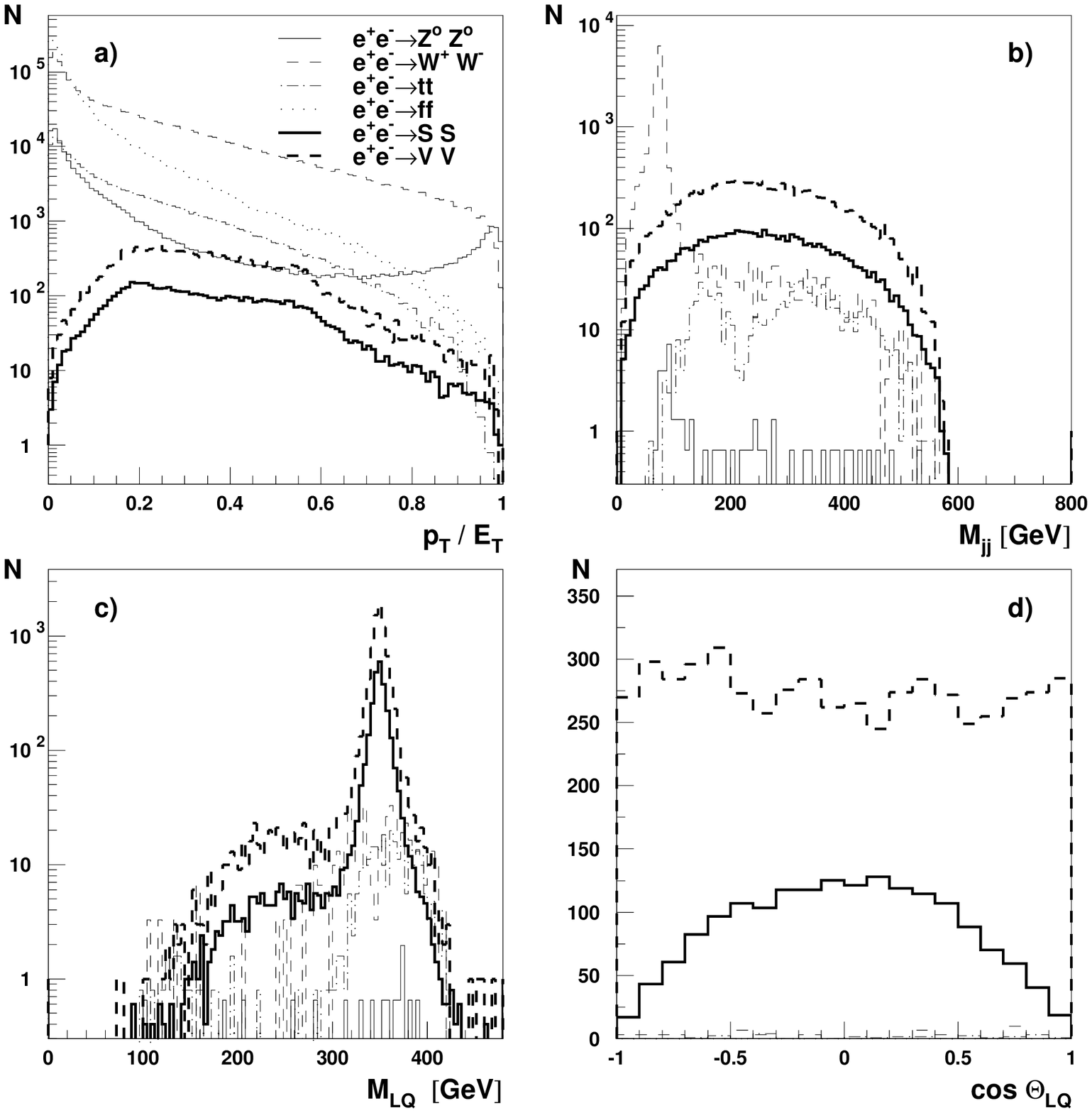}
}}
  \caption{Selection of the leptoquark pair-production events
           in the $e\nu jj$ channel. Distributions for signal 
           (scalar $S_{1/2}$ or vector $V_{\circ}$
            leptoquarks with M=350 GeV) and background events 
           ($Z^{\circ}$ pair-production, $W^{\pm}$ pair-production,
            $t \bar{t}$ pair-production and production of other
            fermion pairs $f \bar{f}$ ($f \ne t$), as indicated in the plot),
            expected for the integrated luminosity of 500$fb^{-1}$, 
            are shown for:
ratio of the total transverse momentum to the transverse energy, before
any selection cuts (a), 
invariant mass  of the jet system $M_{jj}$, 
after pre-selection and lepton identification cuts (b),
reconstructed leptoquark mass $M_{LQ}$, after all selection cuts (c),
leptoquark production angle $\Theta_{LQ}$, in $\pm\;20$ GeV mass window (d).
}
  \label{fig-cuts2}
\end{figure}

Expected signal and background event distributions in the main selection 
variables and in the reconstructed leptoquark production angle
are presented in Figure \ref{fig-cuts2}.
The selection efficiency for scalar and vector leptoquark
pair-production, for leptoquark mass $M_{LQ} = 350$ GeV and
$\sqrt{s} = 800$ GeV, is about 30\%. 
Estimated background from $W^{\pm}$ and $t\bar{t}$ pair-production
corresponds to the cross-section of about 0.13 $fb$
(65 events for 500 $fb^{-1}$) and is taken into account in the presented
analysis.

\subsection{Selection of $ej(j)$ events}

Leptoquarks can be also single-produced at TESLA, in $e^{\pm} \gamma$
collisions:
\begin{eqnarray*}
\gamma e^{-} \;  \rightarrow & LQ + X & \rightarrow \; ej + X
\end{eqnarray*}
where $X$ is the ``photon remnant'': hadronic state remaining
after leptoquark production in electron-quark fusion.
Reconstruction of the remnant jet turns out to be very important for
efficient leptoquark selection and background reduction.
Direction of the remnant production gives an independent tag on the
charge of the scattered $e^{\pm}$, strongly suppressing background from
NC $e^{\pm}\gamma$ DIS event with wrong reconstruction 
of the scattered lepton charge.

Cuts used to select single leptoquark production events:
    
\begin{itemize}

\item Pre-selection cuts:
      \begin{itemize}
      \item total energy $E$ of an event: $E > 0.5 \sqrt{s}$,
      \item total transverse energy $E_{t}$ of an event: 
                               $E_{t} > 50$ GeV,
      \item total transverse momentum $p_{t}$ to $E_{t}$ ratio:
                              $p_{t}/E_{t} < 0.1$,
      \item total invariant mass  $M_{tot}$ of  an event: 
                       $300 < M_{tot} < 600$ GeV,
      \end{itemize}

\item Three energy clusters are reconstructed using the Durham jet algorithm
      (without constraining $y_{cut}$). 
      These clusters are identified as:
      \begin{itemize}
      \item an isolated electron or positron 
            with transverse momentum $p^{e}_{t} > 80$ GeV,
      \item hadronic jet from leptoquark decay
            with transverse momentum $p^{j}_{t} > 80$ GeV,
      \item photon remnant 
            with transverse momentum $p^{r}_{t} < 50$ GeV
            and $p^{r}_{t} < 0.1 \; p^{j}_{t}$; remnant emission
            angle w.r.t. the photon direction $\cos \theta > 0.8$.
      \end{itemize}

\item  mass reconstructed from decay electron energy and 
       direction\footnote{For high electron scattering angles,
       leptoquark mass resolution for ``electron'' method 
       (from electron energy and scattering angle only;
       using incident beam energy) is better than from electron-jet 
       invariant mass (about 4 GeV and 6 GeV respectively).}
       $M_{e}$ within $\pm \; 8$ GeV from the nominal value.
    
\end{itemize}

\begin{figure}[tbp]
\centerline{\resizebox{!}{\figheight}{%
  \includegraphics{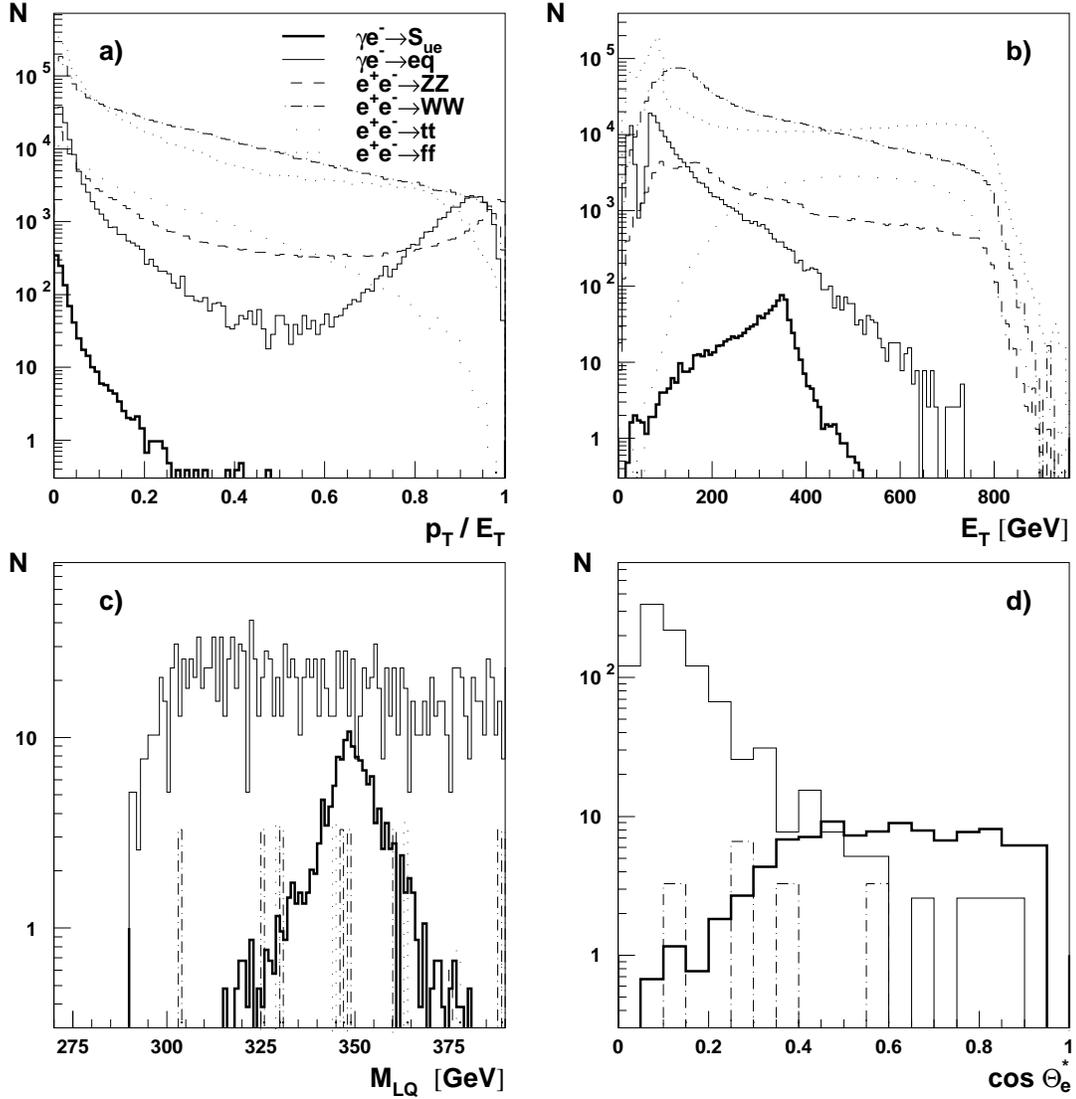}
}}
  \caption{Selection of the single leptoquark production events
           ($ej(j)$ channel). Distributions for signal 
           (scalar $S_{\circ}^{R}$ leptoquark with M=350 GeV
            and $\lambda_{R} = 0.1 \; e$) and background events 
           ($e^{\pm}\gamma$ NC DIS with electron-quark invariant mass
            $M_{eq} > 300$ GeV, 
            $Z^{\circ}$ pair-production, $W^{\pm}$ pair-production,
            $t \bar{t}$ pair-production and production of other
            fermion pairs $f \bar{f}$ ($f \ne t$), as indicated in the plot),
     expected for the integrated $e^{+}e^{-}$ luminosity of 500$fb^{-1}$, 
            are shown for:
ratio of the total transverse momentum to the transverse energy, before
any selection cuts (a), 
the total transverse energy, before any selection cuts (b), 
reconstructed leptoquark mass $M_{LQ}$, after all selection cuts (c),
electron   emission angle in the leptoquark 
rest frame $\theta^{\star}_{e}$, in $\pm\;8$ GeV mass window (d).
}
  \label{fig-cuts3}
\end{figure}
    
Expected signal and background event distributions in the selected
variables are presented in Figure \ref{fig-cuts3}.
    The selection efficiency for scalar leptoquark
production, for leptoquark mass $M_{LQ} = 350$ GeV and
$\sqrt{s} = 800$ GeV, is about 15\%. 
Large, irreducible background comes from $e^{\pm}\gamma$ NC DIS.
However, NC DIS background results in the steeply falling
distribution of the electron emission angle in the leptoquark 
rest frame $\theta^{\star}_{e}$, whereas for scalar leptoquark
production the distribution is flat at high $\theta^{\star}_{e}$
(see Figure \ref{fig-cuts3}d).
Estimated background from remaining processes is about 8 event
(in the selected mass window), mainly from $W$-pair production,
 and is taken into account.


\section{Determination of the Yukawa couplings}
\label{sec-res}

Considered in this note is the possibility of the leptoquark 
Yukawa coupling measurement from the observed angular leptoquark
distributions. Three distributions are considered: 
leptoquark production angle distribution in the pair-production process,
in $eejj$ channel and in $e\nu jj$ channel,
and the decay electron emission angle in the single leptoquark
production process ($ej(j)$ channel). 
Yukawa coupling determination has been studied for three leptoquark
types: $S_{\circ}$ (iso-singlet), $S_{1/2}$ (iso-doublet) and
$S_{1}$ (iso-triplet). 
For $S_{\circ}$  and  $S_{1/2}$ leptoquarks 
both left-handed and right-handed Yukawa couplings 
($\lambda_{L}$ and $\lambda_{R}$) are considered,
whereas $S_{1}$ leptoquark couples only to left-handed leptons.

\begin{figure}[p]
\centerline{\resizebox{!}{6cm}{%
  \includegraphics{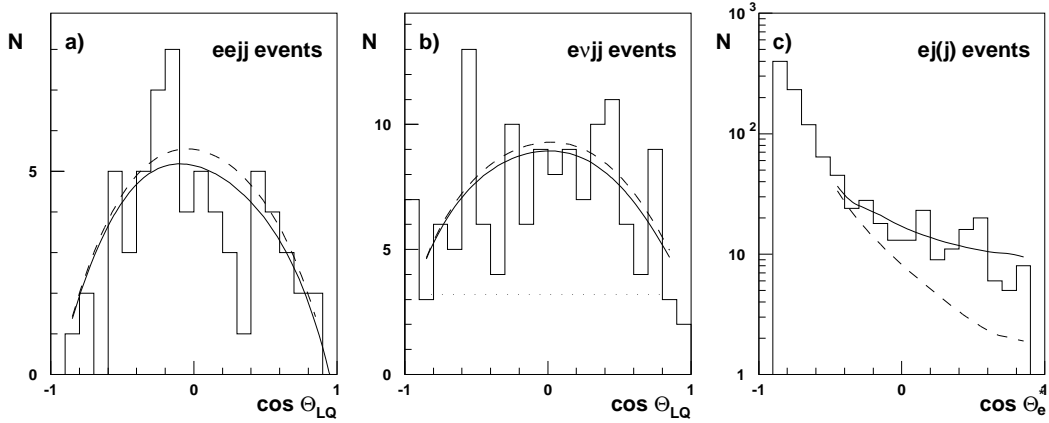}
}}\vspace{-0.7cm}
  \caption{Angular distributions used to determine Yukawa couplings
           of the $S_{\circ}^{L}$ leptoquark: leptoquark
           production angle in the pair-production process, 
           reconstructed in the $eejj$ channel (a), 
           leptoquark production angle
           reconstructed in the $e\nu jj$ channel (b)
           and the decay electron emission angle 
           in the leptoquark rest frame, for the single leptoquark
           production process (c).
     Leptoquark production events were generated assuming
     $M_{LQ} = $350GeV and $\lambda_{L}$=0.15$e$ ($\lambda_{R}$=0).
     Solid lines represent the result of the simultaneous 
     log-likelihood fit to the three distributions.
     Dashed line indicates model expectations for $\lambda_{L}\rightarrow 0$.
     Dotted line shows the expected background from $W^{\pm}$ and
     $t\bar{t}$ pair production (b).
           }
  \label{fig-sc1}
\end{figure}
\begin{figure}[p]
\vspace{-1cm}
\centerline{\resizebox{!}{10cm}{%
  \includegraphics{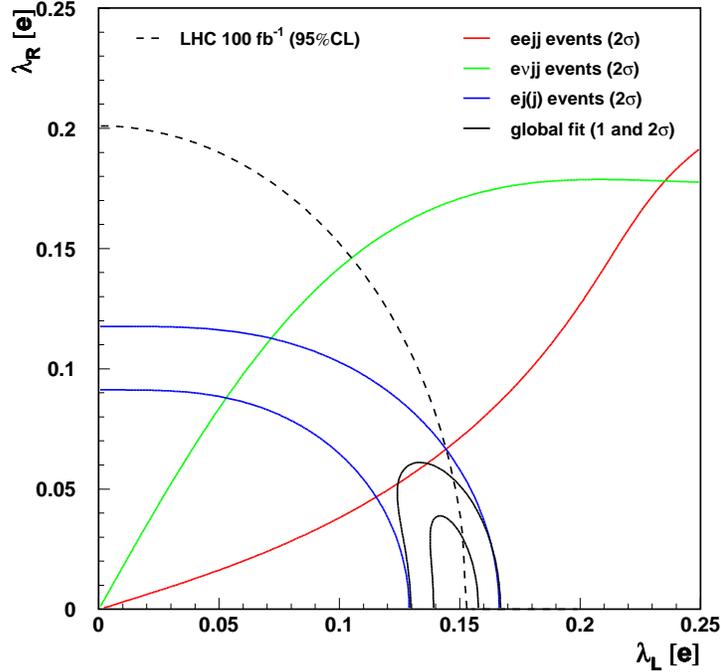}
}}\vspace{-0.7cm}
  \caption{Results of the log-likelihood fit to the 
           $S_{\circ}^{L}$ leptoquark angular distributions shown
           in Figure \ref{fig-sc1}. 
           2$\sigma$ ($\Delta \ln {\cal L}=2$) contours in
           $(\lambda_{L}, \lambda_{R})$, resulting
           from the fits to the separate distributions (as indicated
           in the plot) are compared with
           1$\sigma$ ($\Delta \ln {\cal L}=0.5$) and 2$\sigma$ contours
           from the simultaneous fit to the three distributions.
           Also shown are  95\% CL exclusion limits 
           expected from the Drell-Yan $e^{+}e^{-}$ pair-production
           at the LHC.
           Leptoquark production events were generated assuming
           $M_{LQ} = $350GeV and $\lambda_{L}$=0.15$e$ ($\lambda_{R}$=0).
           Luminosity uncertainty is 1\%.
           }\vspace{-1cm}
  \label{fig-sc2}
\end{figure}

Observed angular distributions, in three considered channels,
for $S_{\circ}^{L}$ leptoquark production are presented 
in Figure \ref{fig-sc1}.
Events were generated assuming  $M_{LQ} = $350GeV and
$\lambda_{L}$=0.15$e$ ($\lambda_{R}$=0).
Numbers of observed events shown on the plots correspond to the 
integrated luminosity of 500 fb$^{-1}$.
Result of the simultaneous log-likelihood fit 
to the three distributions
is indicated by the solid line.
Precision of the coupling determination, for the assumed parameter values
is about 6\%.
This high precision results mainly from the measurement of the single 
leptoquark production process. 
However, single leptoquark production does not
distinguish between left- and right-handed Yukawa couplings.
Measurement of other distributions, related to the leptoquark 
pair-production, is required to verify the chirality of the coupling.
This is demonstrated in Figure \ref{fig-sc2}.
2$\sigma$ ($\Delta \ln {\cal L}=2$) contours in
$(\lambda_{L}, \lambda_{R})$, resulting
from the two parameter fits to the separate distributions  
are compared with 1$\sigma$ ($\Delta \ln {\cal L}=0.5$) 
and 2$\sigma$ contours from the simultaneous fit to 
the three distributions, as show in  Figure \ref{fig-sc1}.
Using information from both single leptoquark production and pair-production
$\lambda_{L}$ and $\lambda_{R}$ can be simultaneously constrained.
The value of the leptoquark Yukawa coupling can be precisely determined
and its chirality unambiguously defined.

\begin{figure}[p]
\centerline{\resizebox{!}{6cm}{%
  \includegraphics{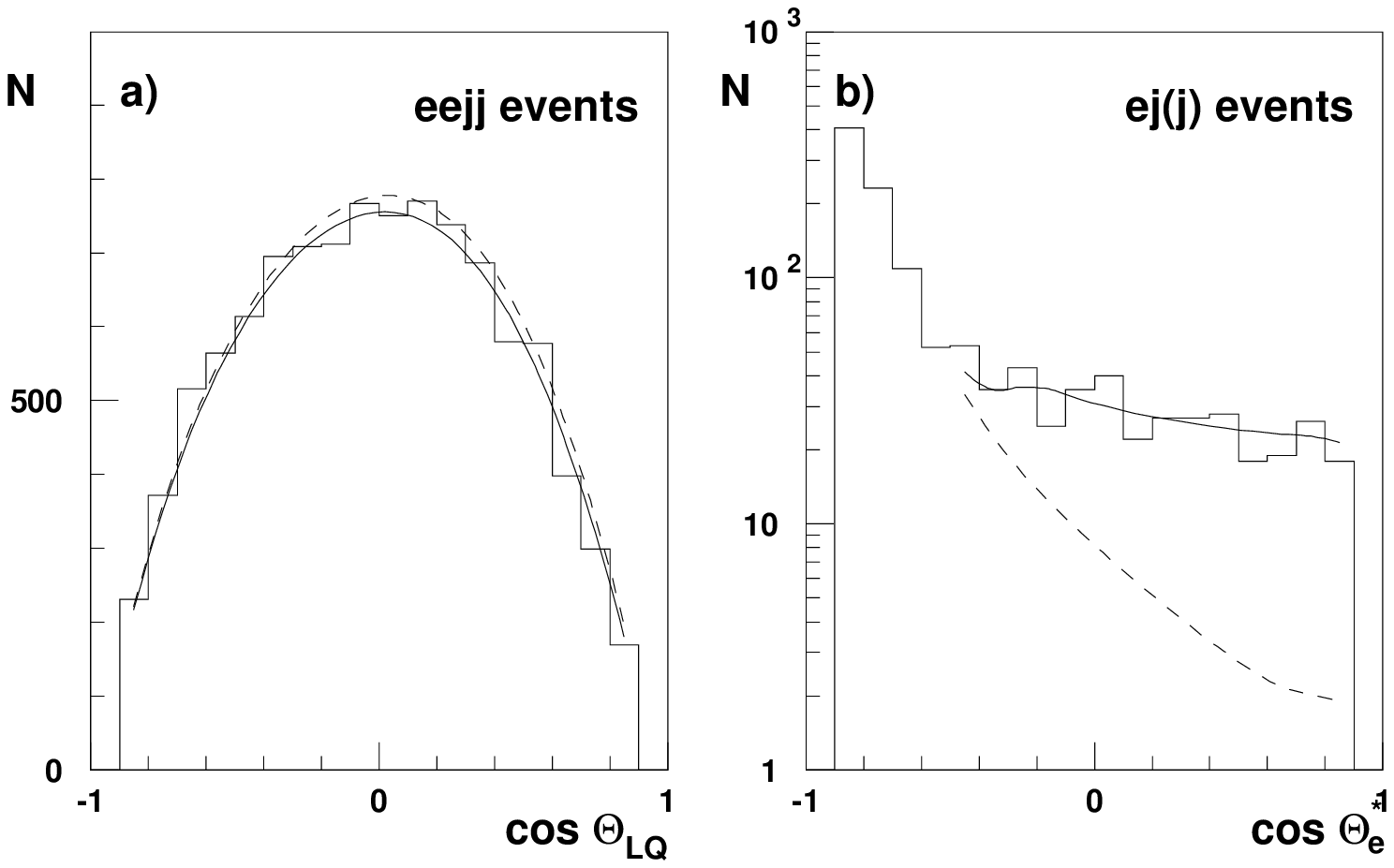}
}}\vspace{-0.7cm}
  \caption{Angular distributions used to determine Yukawa couplings
           of the $S_{1/2}^{R}$ leptoquark: leptoquark
           production angle in the pair-production process, 
           reconstructed in the $eejj$ channel (a)
           and the decay electron emission angle 
           in the leptoquark rest frame, for the single leptoquark
           production process (b).
     Leptoquark production events were generated assuming
     $M_{LQ} = $350GeV and $\lambda_{R}$=0.15$e$ ($\lambda_{L}$=0).
     Solid lines represent the result of the simultaneous 
     log-likelihood fit to the three distributions.
     Dashed line indicates model expectations for $\lambda_{R}\rightarrow 0$.
           }
  \label{fig-sab1}
\end{figure}
\begin{figure}[p]
\vspace{-1cm}
\centerline{\resizebox{!}{10cm}{%
  \includegraphics{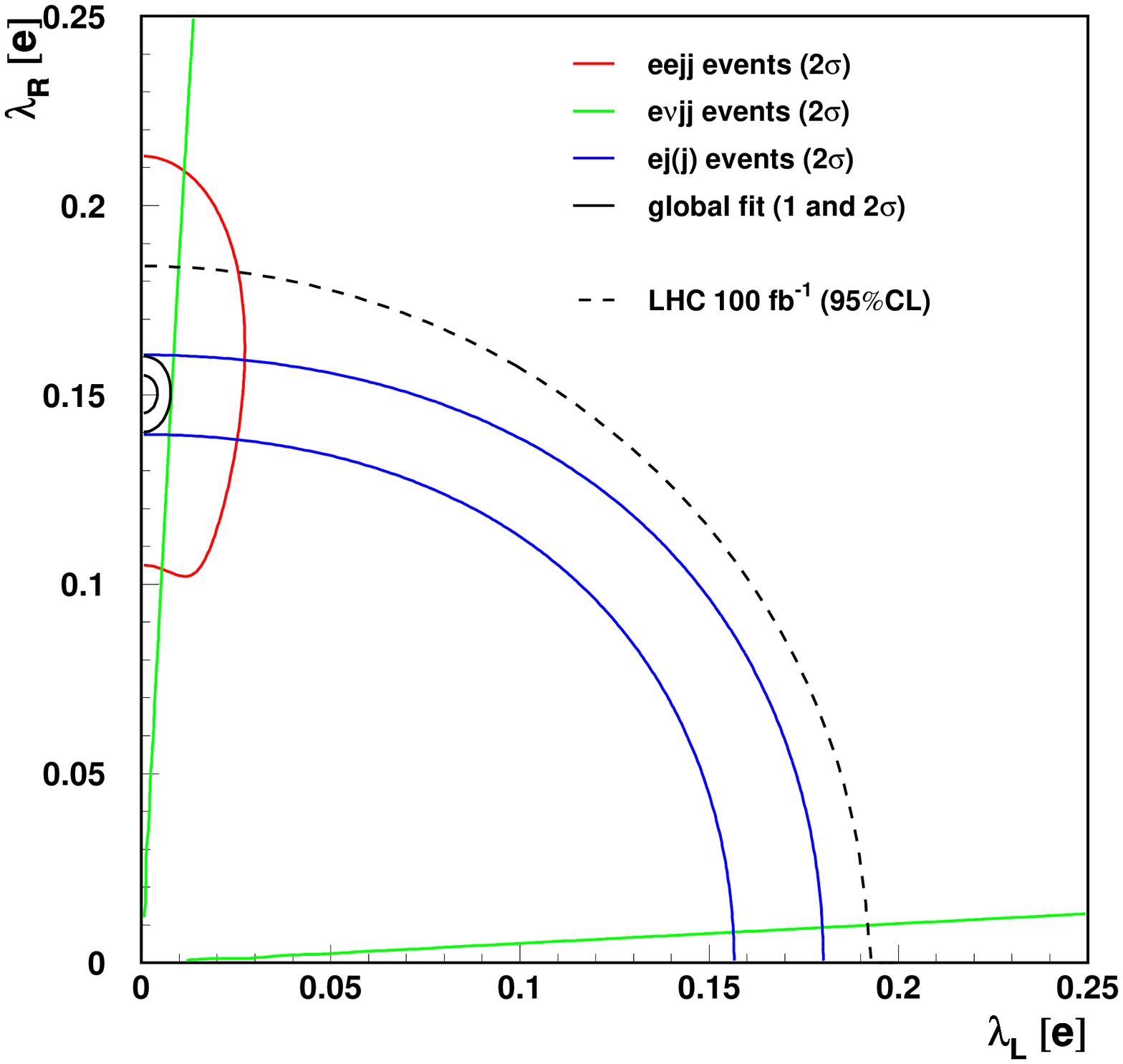}
}}\vspace{-0.7cm}
  \caption{Results of the log-likelihood fit to the 
           $S_{1/2}^{R}$ leptoquark angular distributions shown
           in Figure \ref{fig-sab1}. 
           2$\sigma$ ($\Delta \ln {\cal L}=2$) contours in
           $(\lambda_{L}, \lambda_{R})$, resulting
           from the fits to the separate distributions (as indicated
           in the plot) are compared with
           1$\sigma$ ($\Delta \ln {\cal L}=0.5$) and 2$\sigma$ contours
           from the simultaneous fit to the three distributions.
           Also shown are  95\% CL exclusion limits 
           expected from the Drell-Yan $e^{+}e^{-}$ pair-production
           at the LHC.
           Leptoquark production events were generated assuming
           $M_{LQ} = $350GeV and $\lambda_{R}$=0.15$e$ ($\lambda_{L}$=0).
           Luminosity uncertainty is 1\%.
           }
  \label{fig-sab2}
\vspace{-1cm}
\end{figure}

\begin{figure}[p]
\centerline{\resizebox{!}{6cm}{%
  \includegraphics{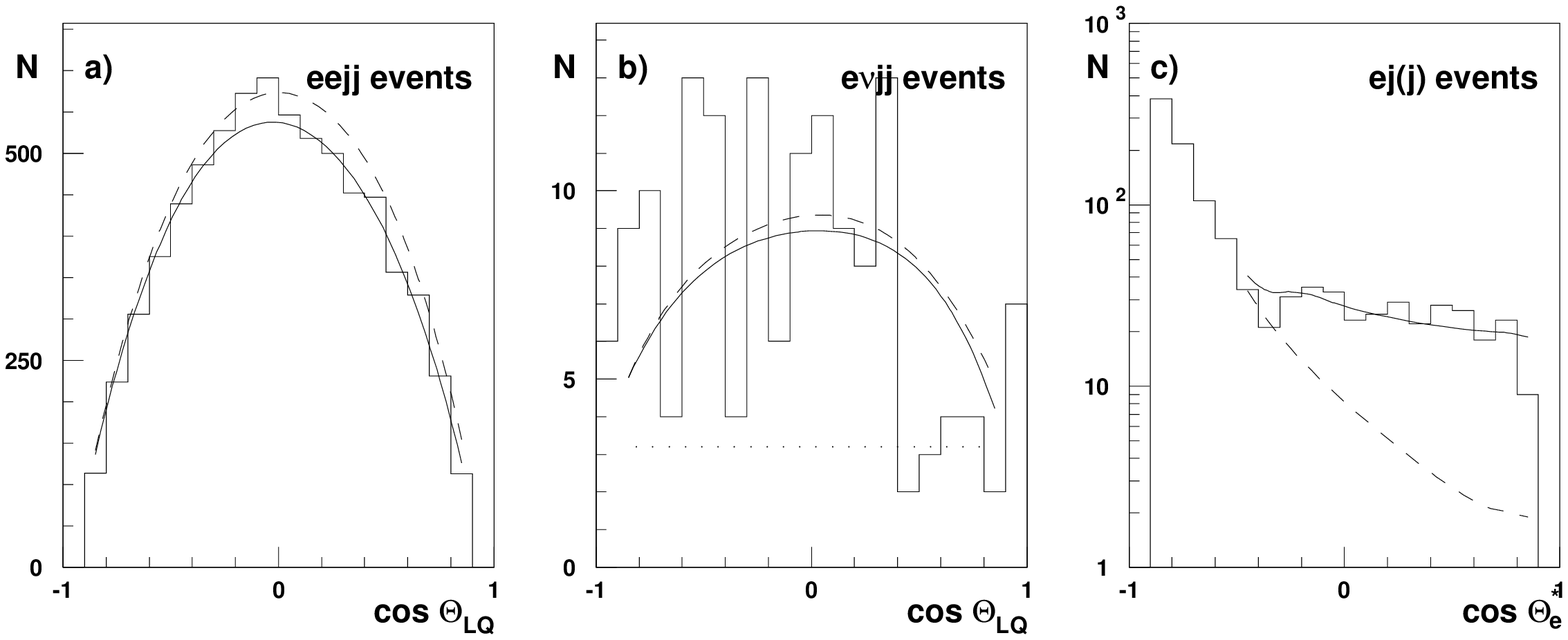}
}}\vspace{-0.7cm}
  \caption{Angular distributions used to determine Yukawa couplings
           of the $S_{1}$ leptoquark: leptoquark
           production angle in the pair-production process, 
           reconstructed in the $eejj$ channel (a), 
           leptoquark production angle
           reconstructed in the $e\nu jj$ channel (b)
           and the decay electron emission angle 
           in the leptoquark rest frame, for the single leptoquark
           production process (c).
     Leptoquark production events were generated assuming
     $M_{LQ} = $350GeV and $\lambda_{L}$=0.15$e$.
     Solid lines represent the result of the simultaneous 
     log-likelihood fit to the three distributions.
     Dashed line indicates model expectations for $\lambda_{L}\rightarrow 0$.
     Dotted line indicates the expected background from $W^{\pm}$ and
     $t\bar{t}$ pair production (b).
           }
  \label{fig-sde1}
\end{figure}

\begin{figure}[p]
\vspace{-1cm}
\centerline{\resizebox{!}{10cm}{%
  \includegraphics{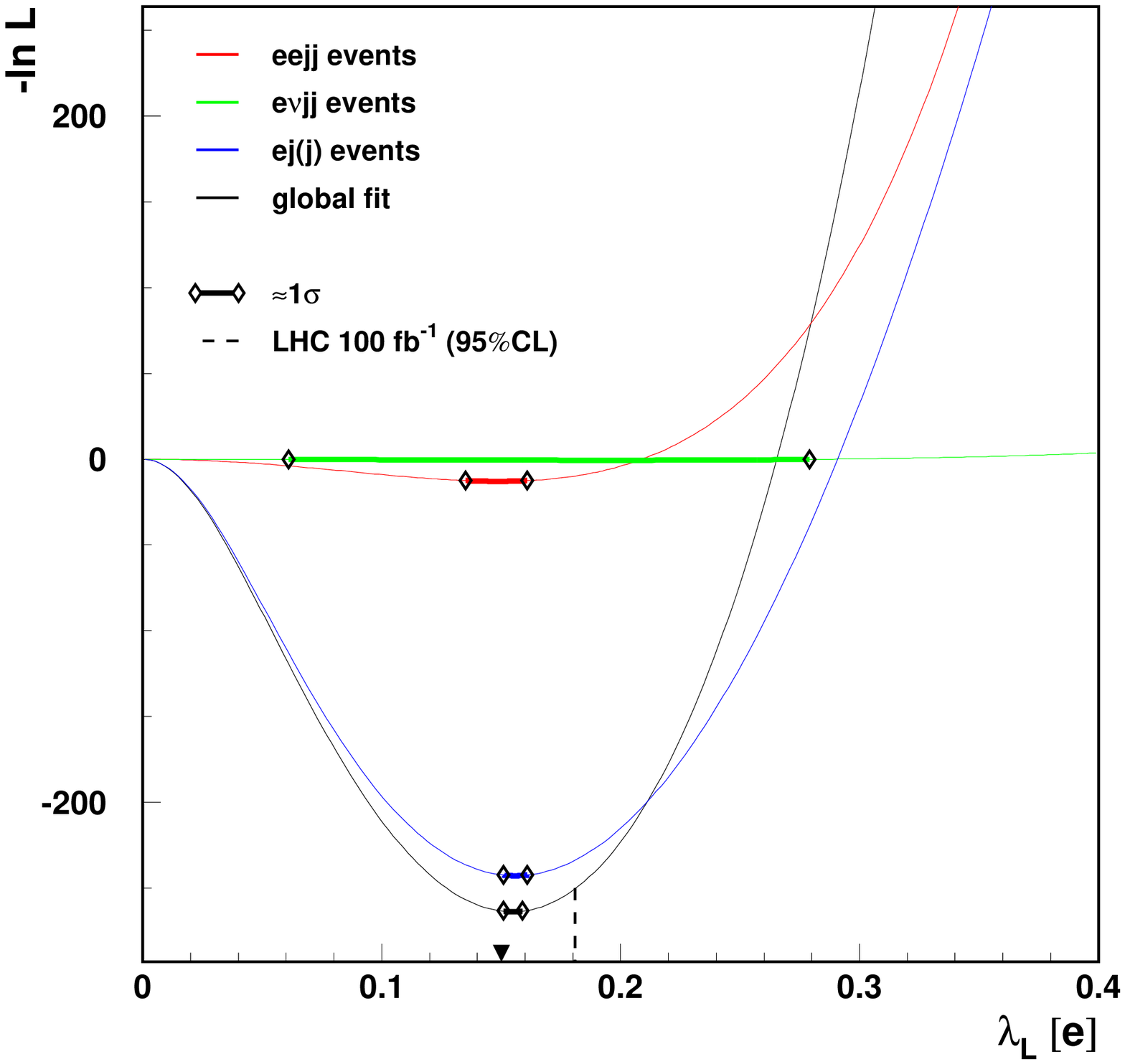}
}}\vspace{-0.7cm}
  \caption{Log-likelihood curve ($-\ln {\cal L}$) for $S_{1}$ leptoquark,
           resulting from the fits of separate angular distributions
           and from the simultaneous fit to the three distributions,
           as indicated in the plot.
           Thicker lines indicate $\pm \; 1 \sigma$ 
           ($\Delta \ln {\cal L}=0.5$) errors on $\lambda_{L}$, 
           resulting from the fits.
           Also shown are 95\% CL exclusion limit 
         expected from the Drell-Yan $e^{+}e^{-}$ pair-production
           at the LHC.
           Leptoquark production events were generated assuming
           $M_{LQ} = $350GeV and $\lambda_{L}$=0.15$e$.
           Luminosity uncertainty is 1\%.
           }
  \vspace{-1cm}
  \label{fig-sde2}
\end{figure}

Measurement of the Yukawa couplings for $S_{\circ}$ would be
the most difficult one at TESLA, as $S_{\circ}$  pair-production 
cross-section is the smallest.
Much better measurement is possible for $S_{1/2}$ and $S_{1}$ leptoquarks.
Corresponding results are shown in Figures \ref{fig-sab1} and \ref{fig-sab2}
(for  $S_{1/2}^R$, assuming $\lambda_{R}$=0.15$e$), and 
in Figures \ref{fig-sde1} and \ref{fig-sde2}
(for  $S_{1}$, assuming $\lambda_{L}$=0.15$e$).
Presented results confirm that precise determination 
of the Yukawa couplings at TESLA will be possible even
in the domain, which will not be accessible at 
LHC.\footnote{Coupling values chosen are just below the expected 
exclusion limits from Drell-Yan electron pair-production at LHC,
with 100 fb$^{-1}$ \cite{myfuture}.}


\begin{figure}[t]
\centerline{\resizebox{!}{8cm}{%
  \includegraphics{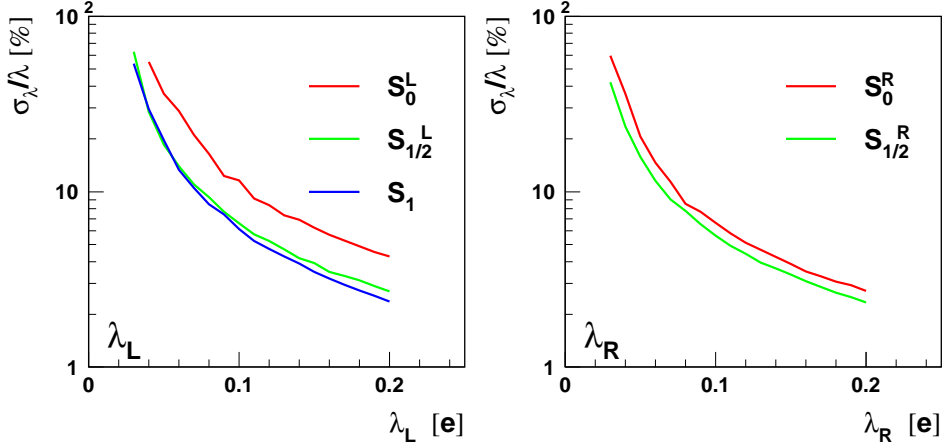}
}}\vspace{-0.7cm}
  \caption{Relative precision of the leptoquark Yukawa 
 coupling determination, for left- (left plot) and right-handed
 (right plot) couplings, for indicated leptoquark models.
  Expected result of the simultaneous log-likelihood fit 
  to the three considered distributions, for $M_{LQ} = $350GeV, as
  a function of the coupling value.
 }
  \label{fig-siglam}
 \vspace{-1cm}
\end{figure}

\begin{figure}[t]
\centerline{\resizebox{!}{8cm}{%
  \includegraphics{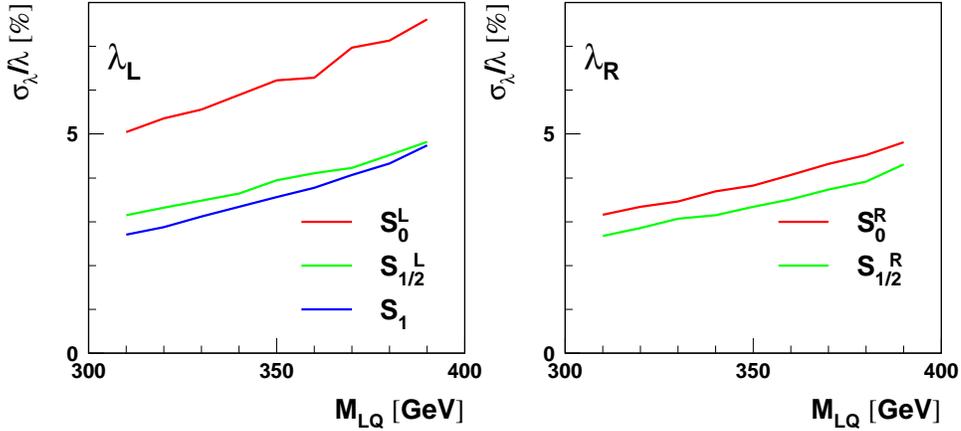}
}}\vspace{-0.7cm}
  \caption{Relative precision of the leptoquark Yukawa 
 coupling determination, for indicated leptoquark models.
  Expected result of the simultaneous log-likelihood fit 
  to the three considered distributions, for $\lambda_{L}$=0.15$e$ (left plot)
  and $\lambda_{R}$=0.15$e$ (right plot), as a function of the leptoquark mass.
 }
  \label{fig-sigmass}
\end{figure}

\begin{figure}[t]
\centerline{\resizebox{!}{10cm}{%
  \includegraphics{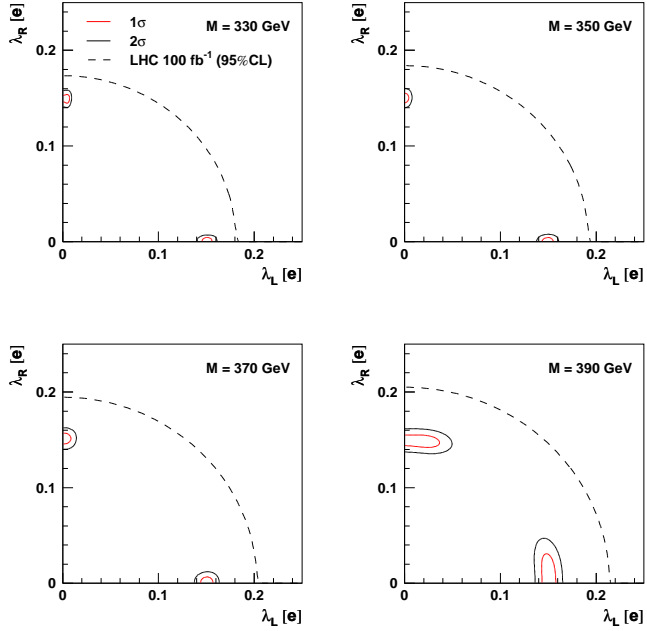}
}}\vspace{-0.5cm}
  \caption{Results of the log-likelihood fit to the 
           $S_{1/2}$ leptoquark angular distributions.
           1$\sigma$ and 2$\sigma$ contours in  $(\lambda_{L}, \lambda_{R})$, 
           resulting from the simultaneous fit to all considered
           distributions are compared for different leptoquark masses, 
           as indicated in the plot.
           Also shown are 95\% CL exclusion limits 
           expected from the Drell-Yan $e^{+}e^{-}$ pair-production
           at the LHC.
           Leptoquark production events were generated assuming
           $\lambda_{L}$=0.15$e$ or $\lambda_{R}$=0.15$e$.
           Luminosity uncertainty is 1\%.
           }
  \label{fig-sab3}
\end{figure}

Precision of the leptoquark Yukawa coupling determination, for different
scalar leptoquark models, has been also studied as a function of the coupling
value (for leptoquark mass $M_{LQ}$=350 GeV) and as a function 
of the leptoquark mass (for fixed value of coupling: 
$\lambda_{L}$=0.15$e$ or $\lambda_{R}$=0.15$e$).
Results are shown in Figures \ref{fig-siglam} and \ref{fig-sigmass}
respectively.
Determination of the leptoquark Yukawa coupling,
seems to be possible down to the coupling values or about 0.05$e$.
Precision of the measurement very slowly deteriorates with the increasing 
leptoquark mass (for masses up to 390 GeV), as it results mainly from the
single leptoquark production measurements.
Shown in Figures \ref{fig-sab3} are 
1$\sigma$ and 2$\sigma$ contours in  $(\lambda_{L}, \lambda_{R})$, 
for $S_{1/2}$ leptoquark with $\lambda_{L}$=0.15$e$ or $\lambda_{R}$=0.15$e$,
for leptoquark mass $M_{LQ}=$330, 350, 370 and 390 GeV.
Measurement of the leptoquark pair-production process will
allow to distinguish between left- and right-handed Yukawa couplings
even for leptoquark masses very close to the pair-production kinematic 
limit.


%
\section{Summary}

Measurement of the Yukawa couplings of the first-generation leptoquarks
has been studied for $e^{+}e^{-}$ collisions at TESLA, at $\sqrt{s}=$800 GeV.
By combining measurements from different production and decay channels,
determination of Yukawa couplings with precision on the few per-cent level
is possible.
TESLA will be sensitive to very small leptoquark Yukawa couplings
not accessible at LHC, down to $\lambda_{L,R} \sim 0.05 e$.
Distinction between left-handed and right-handed Yukawa 
couplings is feasible even for leptoquark masses very close to the
pair-production kinematic limit.

TESLA sensitivity to very small Yukawa couplings is limited mainly
by the statistics of single leptoquark production events.
In $e^{+}e^{-}$ running mode of TESLA, considered in this paper, 
single leptoquark production is
suppressed by a very small flux of high energy 
bremsstrahlung and beamstrahlung photons.
If TESLA collider is run in $e\gamma$ mode, cross-section for single
leptoquark production will increase significantly
and much smaller Yukawa couplings can be searched for.

%
%


\end{document}